\documentclass[12pt]{article}

\usepackage{amsthm}
\theoremstyle{plain}

\theoremstyle{definition}

\theoremstyle{proposition}

\theoremstyle{lemma}

\theoremstyle{remark}

\usepackage{amsmath}
\usepackage{amssymb}
\usepackage[dvipdfmx]{graphicx}
\usepackage{bm}

\usepackage[bookmarksnumbered=true,colorlinks=true,filecolor=blue,linkcolor=blue,urlcolor=blue,citecolor=red,linktocpage=true]{hyperref}

\makeatletter
 
  \@addtoreset{equation}{section}
 \makeatother

\newcommand{\be}{\begin{equation}}
\newcommand{\ee}{\end{equation}}
\newcommand{\ba}{\begin{aligned}}
\newcommand{\ea}{\end{aligned}}

\begin{document}
\setlength{\oddsidemargin}{0cm}
\setlength{\baselineskip}{7mm}

\begin{titlepage}
\begin{flushright}   \end{flushright} 

~~\\

\vspace*{0cm}
    \begin{large}
       \begin{center}
         {T-symmetry in String Geometry Theory}
       \end{center}
    \end{large}
\vspace{1cm}

\begin{center}
Matsuo Sato$^{\dagger}$\footnote
           {
e-mail address : msato@hirosaki-u.ac.jp}
 and Taiki Tohshima$^{\dagger}$\footnote
           {e-mail address : h20ms117@hirosaki-u.ac.jp}

\vspace{1cm}

$^{\dagger}$ \it{Department of Mathematics and Physics, Graduate School of Science and Technology, Hirosaki University, Bunkyo-cho 3, Hirosaki, Aomori 036-8561, Japan}
\end{center}

\hspace{5cm}

\begin{abstract}
\noindent
String geometry theory is one of the candidates of non-perturbative formulation of string theory.  In this paper,  we have shown that  dimensionally reduced string geometry theories have what we call T-symmetry. In case of the dimensional reduction in space-like directions, the T-symmetry transformation gives the T-dual transformation between the type IIA and IIB perturbative vacua. In case of  the dimensional reduction in the direction of string geometry time $\bar{\tau}$,  the T-symmetry transformation is independent of the T-dual transformation, and gives a symmetry that cannot be seen in the perturbative string theories.

\end{abstract}

\vfill
\end{titlepage}
\vfil\eject

\setcounter{footnote}{0}

\section{Introduction}
\setcounter{equation}{0}

Perturbative string theories are defined on various perturbatively  stable vacua, which are connected by dualities. It is thought that all the perturbatively stable vacua are included in a non-perturbative formulation of string theory, and are connected by its symmetries and dualities. Concerning T-duality \cite{Kikkawa:1984cp, Buscher:1987sk, Duff:1989tf}, double field theory was proposed in \cite{Siegel:1993xq, Siegel:1993th, Hull:2009mi, Hohm:2010pp} and has been intensively studied as a simplified model of only massless modes in string theory that includes both the type IIA and IIB vacua. However, we should study how dualities are realized among perturbatively stable vacua not only in such a simplified model but also in non-perturbative formulations of string theory themselves. 
 
 From string geometry theory, which is a candidate of the non-perturbative formulation of string theory,  one can derive the path-integrals of  the type IIA , IIB, SO(32) type I, and SO(32) and $E_8 \times E_8$ heterotic perturbative string theories in curved backgrounds \cite{Sato:2017qhj, Sato:2019cno, Sato:2020szq, Sato:2022owj, Sato:2022brv}. That is, the  single string geometry theory includes these perturbatively stable vacua. One can also obtain type IIA , IIB, SO(32) type I, and SO(32) and $E_8 \times E_8$ heterotic supergravities by consistent truncations from the string geometry theory \cite{Honda:2020sbl, Honda:2021rcd}. That is, string geometry theory includes string backgrounds not as external fields like the perturbative string theories. Dynamics of string backgrounds are a part of  dynamics of  the fields in the theory. 

 In this paper,   we investigate how dualities are realized among perturbatively stable vacua in string geometry theory. As a result, we find that T-duality between the type IIA and IIB supergravities are generalized to what we call T-symmetry in dimensionally reduced string geometry theory.

 The organization of the paper is as follows. In section 2, we minimally review string geometry theory.  In section 3, we show that dimensionally reduced string geometry actions have T-symmetry, which is a natural generalization of T-duality between the perturbative vacua.  In section 4, we conclude and discuss our results. 
\vspace{1cm}

\section{Review on string geometry}
\setcounter{equation}{0}

String manifold is constructed by patching open sets in string model space $E$, whose definition is given in \cite{Sato:2017qhj} and summarized as follows. 
First, one of the coordinates is spanned by string geometry time $\bar{\tau} \in \mathbb{R}$ and another is spanned by super Riemann surfaces $\bar{\bold{\Sigma}} \in \mathcal{M}$, where $\mathcal{M}$ is a moduli space of the Riemannian surfaces. 
On each super Riemann surface $\bar{\bold{\Sigma}}$,  
 a global time is defined canonically and uniquely by the real part of the integral of an Abelian differential  \cite{Krichever:1987a, Krichever:1987b}.
We identify this global time as $\bar{\tau}$ and restrict $\bar{\bold{\Sigma}}$ to a $\bar{\tau}$ constant line, and obtain $\bar{\bold{\Sigma}}|_{\bar{\tau}}$. 
An embedding of $\bar{\bold{\Sigma}}|_{\bar{\tau}}$ to $\mathbb{R}^{1, d-1}$ represents a many-body state of superstrings in $\mathbb{R}^{1, d-1}$, and is parametrized by the other coordinates $ \bold{X}_{\hat{D}_{T}}(\bar{\tau})$ where $\hat{D}_T$ represents all the backgrounds except for the target metric, that consist of the B-field, the dilaton and the R-R fields.

Here, we define a string state as an equivalence class $[\bar{\bold{\Sigma}},  \bold{X}_{\hat{D}_{T}}(\bar{\tau}\simeq \pm \infty), \bar{\tau}\simeq \pm \infty]$ by a relation $(\bar{\bold{\Sigma}},  \bold{X}_{\hat{D}_{T}}(\bar{\tau}\simeq \pm \infty), \bar{\tau}\simeq \pm \infty) \sim (\bar{\bold{\Sigma}}',  \bold{X}'_{\hat{D}_{T}}(\bar{\tau}\simeq \pm \infty), \bar{\tau}\simeq \pm \infty)$ if   $\bar{\bold{\Sigma}}|_{\bar{\tau}\simeq \pm \infty}=\bar{\bold{\Sigma}}'|_{\bar{\tau}\simeq \pm \infty}$ and $ \bold{X}_{\hat{D}_{T}}(\bar{\tau}\simeq \pm \infty)= \bold{X}'_{\hat{D}_{T}}(\bar{\tau}\simeq \pm \infty)$. 
String model space $E$  is defined by the collection of the string states $[\bar{\bold{E}}, \bold{X}_{\hat{D}_{T}}(\bar{\tau}), \bar{\tau}]$ by considering all the  $\bar{\bold{\Sigma}}$, all the values of $\bar{\tau}$, and all the $\bold{X}_{\hat{D}_{T}}(\bar{\tau})$. 
$\bar{\bold{E}}$ is a super vierbein on  $\bar{\bold{\Sigma}}$ where $\,\, \bar{}$ represents a representative of the super diffeomorphism and super Weyl transformation on the worldsheet. Giving a super Riemann surface $\bar{\bold{\Sigma}}$ is equivalent to giving a  supervierbein $\bar{\bold{E}}$ up to super diffeomorphism and super Weyl transformations.

How near the two string states is defined by how near the values of $\bar{\tau}$ and how near $\bold{X}_{\hat{D}_{T}}(\bar{\tau})$.
An $\epsilon$-open neighborhood of $[\bar{{\boldsymbol \Sigma}}, \bold{X}_{\hat{D}_{T}s}(\bar{\tau}_s), \bar{\tau}_s]$ is defined by
\begin{eqnarray}
U([\bar{\bold{E}}, \bold{X}_{\hat{D}_{T}s}(\bar{\tau}_s), \bar{\tau}_s], \epsilon)
:=
\left\{[\bar{\bold{E}}, \bold{X}_{\hat{D}_{T}}(\bar{\tau}), \bar{\tau}] \bigm|
\sqrt{|\bar{\tau}-\bar{\tau}_s|^2
+\| \bold{X}_{\hat{D}_{T}}(\bar{\tau}) -\bold{X}_{\hat{D}_{T} s}(\bar{\tau}_s) \|^2}
<\epsilon   \right\}, \label{SuperNeighbour}
\end{eqnarray}
where $\bar{\bold{E}}$  is a discrete variable in the topology of string geometry.
The precise definition of the string topology is given in the section 2 in \cite{Sato:2017qhj}. 

By this definition, arbitrary two string states on a connected super Riemann surface in $E$ are connected continuously. Thus, there is an one-to-one correspondence between a super Riemann surface in $\mathbb{R}^{1, d-1}$ and a curve  parametrized by $\bar{\tau}$ from $\bar{\tau}=-\infty$ to $\bar{\tau}=\infty$ on $E$. That is, curves that represent asymptotic processes on $E$ reproduce the right moduli space of the super Riemann surfaces in $\mathbb{R}^{1, d-1}$. Therefore, string geometry theory possesses all-order information of superstring theory.  Actually, 
the path-integrals of  the perturbative superstring theories in curved backgrounds
are derived from the string geometry theory as in  
\cite{Sato:2017qhj, Sato:2019cno, Sato:2020szq, Sato:2022owj, Sato:2022brv}. The consistency of the perturbation theory determines $d=10$ (the critical dimension). 

A Riemannian string manifold is defined by introducing a metric on a string manifold. First, the cotangent space is spanned by 
\begin{eqnarray}
d \bold{X}^{10}_{\hat{D}_{T}} &:=& d \bar{\tau}
\nonumber \\
d \bold{X}^{(\mu \bar{\sigma} \bar{\theta}) }_{\hat{D}_{T}}&:=& d \bold{X}^{\mu}_{\hat{D}_{T}} \left( \bar{\sigma}, \bar{\tau}, \bar{\theta} \right), \label{cotangen}
\end{eqnarray}
where $d \bar{\bold{E}}$  cannot be a part of basis that span the cotangent space because $\bar{\bold{E}}$  is a discrete variable as in (\ref{SuperNeighbour}). Here, we treat $(\mu \bar{\sigma} \bar{\theta})$ as indices, and $\mu=0, \dots, 9$. 
The summation over $(\bar{\sigma}, \bar{\theta})$ is defined by 
$\int d\bar{\sigma}d^2\bar{\theta} \hat{\bold{E}}(\bar{\sigma}, \bar{\tau}, \bar{\theta})$.
$\hat{\bold{E}}(\bar{\sigma}, \bar{\tau}, \bar{\theta})
:=
\frac{1}{\bar{n}}\bar{\bold{E}}(\bar{\sigma}, \bar{\tau}, \bar{\theta})$, where $\bar{n}$ is the lapse function of the two-dimensional metric.
This summation is transformed as a scalar under $\bar{\tau} \mapsto \bar{\tau}'(\bar{\tau}, \bold{X}_{\hat{D}_T}(\bar{\tau}))$ and invariant under a supersymmetry transformation $(\bar{\sigma}, \bar{\theta}) \mapsto (\bar{\sigma}'(\bar{\sigma}, \bar{\theta}), \bar{\theta}' (\bar{\sigma}, \bar{\theta}))$. 
Because all indices are contracted, the action (\ref{action of bos string-geometric model}) is invariant under this $\mathcal{N}=(1,1)$ supersymmetry transformation.
An explicit form of the line element is given by
\begin{eqnarray}
&&ds^2(\bar{\bold{E}}, \bold{X}_{\hat{D}_{T}}(\bar{\tau}), \bar{\tau}) \nonumber \\
=&&G(\bar{\bold{E}}, \bold{X}_{\hat{D}_{T}}(\bar{\tau}), \bar{\tau})_{10 \; 10} (d\bar{\tau})^2 \nonumber \\ 
&&+2 d\bar{\tau} \int d\bar{\sigma} d^2\bar{\theta}  \hat{\bold{E}} \sum_{\mu} G(\bar{\bold{E}}, \bold{X}_{\hat{D}_{T}}(\bar{\tau}), \bar{\tau})_{10 \; (\mu \bar{\sigma} \bar{\theta})} d \bold{X}_{\hat{D}_{T}}^{ (\mu \bar{\sigma} \bar{\theta})}( \bar{\tau}) \nonumber \\
&&+\int d\bar{\sigma} d^2\bar{\theta} \hat{\bold{E}}  \int d\bar{\sigma}'  d^2\bar{\theta}' \hat{\bold{E}}'  \sum_{\mu, \mu'} G(\bar{\bold{E}}, \bold{X}_{\hat{D}_{T}}(\bar{\tau}), \bar{\tau})_{ \; (\mu \bar{\sigma} \bar{\theta})  \; (\mu' \bar{\sigma}' \bar{\theta}')} d \bold{X}_{\hat{D}_{T}}^{ (\mu \bar{\sigma} \bar{\theta})}( \bar{\tau}) d \bold{X}_{\hat{D}_{T}}^{ (\mu' \bar{\sigma}' \bar{\theta}')}( \bar{\tau}).\nonumber \\
&& \label{LineElement}
\end{eqnarray}
Here, we should note that the fields are functionals of $\bar{\bold{E}}$. The inverse metric $\mathbf{G}^{\bold{I}\bold{J}}(\bar{\bold{E}}, \bold{X}_{\hat{D}_{T}}(\bar{\tau}), \bar{\tau})$\footnote{Like this, the fields $\mathbf{G}_{\bold{I}\bold{J}}$, $\bold{\Phi}$, $\mathbf{B}_{\bold{L}_1\bold{L}_2}$ and $\bold{A}_{\bold{L}_1 \cdots \bold{L}_{p-1}}$ are functionals of the coordinates $\bar{\bold{E}}$, $\bold{X}_{\hat{D}_{T}}(\bar{\tau})$ and $\bar{\tau}$.} is defined by $\mathbf{G}_{\bold{I}\bold{J}}\mathbf{G}^{\bold{J}\bold{K}}=\mathbf{G}^{\bold{K}\bold{J}}\mathbf{G}_{\bold{J}\bold{I}}=\delta^{\bold{K}}_{\bold{I}}$, where $\delta^{10}_{10}=1$ and $\delta_{\mu \bar{\sigma} \bar{\theta}}^{\mu' \bar{\sigma}' \bar{\theta}'}=\delta_{\mu}^{\mu'} \delta_{\bar{\sigma} \bar{\theta}}^{\bar{\sigma}' \bar{\theta}'}$, where $\delta_{\bar{\sigma} \bar{\theta} }^{\bar{\sigma}' \bar{\theta}'}=\delta_{(\bar{\sigma} \bar{\theta}) (\bar{\sigma}' \bar{\theta}')}=\frac{1}{\hat{\bold{E}}}\delta(\bar{\sigma}-\bar{\sigma}') \delta^2(\bar{\theta}-\bar{\theta}')$. 

In the following sections, we will study string geometry theory whose action is given by,
\begin{eqnarray}
S=\int \mathcal{D}\bold{E} \mathcal{D}\bar{\tau} \mathcal{D}\bold{X}_{\hat{D}_T} \sqrt{-\bold{G}} \left( e^{-2 \bold{\Phi}} \left( \mathbf{R}  + 4 \nabla_{\bold{I}} \bold{\Phi} \nabla^{\bold{I}} \bold{\Phi} - \frac{1}{2} |\tilde{\mathbf{H}} |^{2} -\mbox{tr}(|\mathbf{F}|^2)  \right) -\frac{1}{2}\sum_{p=1}^{\infty}
|\tilde{\bold{F}}_p|^2     
\right),
\label{action of bos string-geometric model}
\end{eqnarray}
where we use the Einstein notation for the index $\bold{I}=\{10,(\mu \bar{\sigma} \bar{\theta}) \}$
where
$\mu$, $\bar{\sigma}$, and $\bar{\theta}$ represent the directions of the target space of strings,
a coordinate that specify the strings,
and its super partner, respectively.
The equations of motion of this model can be consistently truncated to the ones of all the ten-dimensional  supergravities, namely type IIA, IIB, SO(32) type I, and SO(32) and $E_8 \times E_8$ heterotic supergravities as in \cite{Honda:2021rcd}. That is, this model includes all the superstring backgrounds.

The action~(\ref{action of bos string-geometric model}) consists of  a scalar curvature    $\mathbf{R}$ of a metric $\mathbf{G}_{\bold{I}_{1} \bold{I}_{2}}$, a scalar field $\bold{\Phi}$, a field strength $\tilde{\mathbf{H}}=d\mathbf{B}-\boldsymbol{\omega}_3$ of a two-form field $\mathbf{B}_{\bold{I}_{1} \bold{I}_{2}}$, where
$\boldsymbol{\omega}_3=\mbox{tr}(\mathbf{A}\wedge d\mathbf{A} -\frac{2i}{3}\mathbf{A} \wedge \mathbf{A} \wedge \mathbf{A})$, and 
$\mathbf{A}$ is a $N \times N$ Hermitian gauge field, whose field strength is given by $\mathbf{F}$, and p-forms $\tilde{\bold{F}}_p$. $\tilde{\bold{F}}_p$ are defined by 
$\sum_{p=1}^{\infty} \tilde{\bold{F}}_p=e^{-\bold{B}_2}\wedge \sum_{k=1}^{\infty} \bold{F}_k$, where $\bold{F}_k$ are field strengths of (k-1)-form fields $\bold{A}_{k-1}$. 
Explicitly, 
$ \tilde{\bold{F}}_p
=\sum_{k=0}^{\lfloor (p-1)/2\rfloor}\frac{1}{k!}(-\bold{B}_2)^k\bold{F}_{p-2k}$.
The relations between $\bold{A}_p$ and the R-R fields $\bold{C}_p$ are given by 
$\bold{A}_p
=\sum_{k=0}^{\lfloor p/2\rfloor}\frac{1}{k!}(\bold{B}_2)^k\wedge\bold{C}_{p-2k}$, and then
 $\tilde{\bold{F}}_p=d\bold{C}_{p-1}+\bold{H}_3\wedge\bold{C}_{p-3}.$

\vspace{1cm}

\section{T-symmetry in string geometry}
\setcounter{equation}{0}
We will show that there are what we call T-symmetries in the actions obtained by performing dimensional reductions in spacial directions in subsection 3.1 and in the temporal direction in subsection 3.2.  In the spacial case, T-symmetry gives T-duality between the perturbative string theories, whereas in the temporal case, T-symmetry is a new symmetry which cannot be seen in the perturbative string theories.    
In the following, we consider the case that the gauge field $\mathbf{A}$ decouples, where the action (\ref{action of bos string-geometric model}) becomes 
\begin{eqnarray}
S'=\int \mathcal{D}\bold{E} \mathcal{D}\bar{\tau} \mathcal{D}\bold{X}_{\hat{D}_T} \sqrt{-\bold{G}} \left( e^{-2 \bold{\Phi}} \left( \mathbf{R}  + 4 \nabla_{\bold{I}} \bold{\Phi} \nabla^{\bold{I}} \bold{\Phi} - \frac{1}{2} |\mathbf{H} |^{2}  \right) -\frac{1}{2}\sum_{p=1}^{\infty}
|\tilde{\bold{F}}_p|^2     
\right),
\label{action of T-symmetry}
\end{eqnarray}
where $\mathbf{H}=d\mathbf{B}$. 

\subsection{T-symmetry in spacial directions}
In this subsection, we consider the action obtained by dimensional reduction in all the $\bold{X}_{\hat{D}_T}^{(9 \bar{\sigma} \bar{\theta})}$ directions of the action (\ref{action of T-symmetry}), where all the fields $\boldsymbol{\varphi}$ satisfy 
\begin{equation}
\partial_{ (9 \bar{\sigma} \bar{\theta})} \boldsymbol{\varphi}=0. \label{DimensinalReduction}
\end{equation} 
We decompose the index of the coordinates as $\bold{M}=(M', 9)$, where $x^9$ is the zero mode of $\bold{X}_{\hat{D}_T}^{(9 \bar{\sigma} \bar{\theta})}$ and $M'$ represents the other directions. We decompose the metric as
\begin{align}
  \bold{G}_{\bold{M} \bold{N}}=\left(
    \begin{array}{cc}
      \bold{g}_{{M'}{N'}}+\bold{k}^2 \bold{A}^{(1)}_{M'} \bold{A}^{(1)}_{N'} & -\bold{k}^2 \bold{A}^{(1)}_{M'} \\
      -\bold{k}^2 \bold{A}^{(1)}_{N'} & \bold{k}^2
    \end{array} \label{decomposition1}
  \right),
\end{align}
where the inverse is given by 
\begin{align}
  \bold{G}^{\bold{M} \bold{N}}=\left(
    \begin{array}{cc}
      \bold{g}^{{M'}{N'}} & \bold{A}^{(1){M'}} \\
      \bold{A}^{(1){N'}} & \bold{k}^{-2}+\bold{g}^{P' Q'}\bold{A}^{(1)}_{P'} \bold{A}^{(1)}_{Q'}
    \end{array}
  \right).
\end{align}
We also decompose the other fields as
\begin{align}
  \bold{\Phi}&= \bold{\phi}+\frac{1}{2}\ln \bold{k} \label{dil}\\ 
  \bold{B}_{{M'} 9}&=\bold{A}^{(2)}_{M'} \\
  \bold{B}_{{M'}{N'}}&=\bold{B}_{{M'}{N'}}+\bold{A}^{(1)}_{[{M'}}\bold{A}^{(2)}_{{N'}]}\\
  \bold{C}_{{M'}_1\cdots{M'}_{2n-1}9}&= -\bold{C}^B_{{M'}_1\cdots{M'}_{2n-1}} \label{eq:cmd} \\
  \bold{C}_{{M'}_1\cdots{M'}_{2n}}&= \bold{C}^B_{{M'}_1\cdots{M'}_{2n}}-2n\bold{A}^{(1)}_{[{M'}_1}\bold{C}^B_{{M'}_2\cdots{M'}_{2n}]} \label{eq:cmn} \\
  \bold{C}_{{M'}_1\cdots{M'}_{2n-2}9}&=\bold{C}^A_{{M'}_1\cdots{M'}_{2n-2}} \label{eq:cmdB} \\
  \bold{C}_{{M'}_1\cdots{M'}_{2n-1}}&= \bold{C}^A_{{M'}_1\cdots{M'}_{2n-1}}-(2n-1)\bold{A}^{(1)}_{[{M'}_1}\bold{C}^A_{{M'}_2\cdots{M'}_{2n-1    }]}. \label{decomposition2}
\end{align}
From these decomposition, the $2n$-form field strengths follow
\begin{eqnarray}
 \tilde{\bold{F}}_{{M'}_1\cdots {M'}_{2n-1}9}&=&\bold{G}^A_{{M'}_1\cdots{M'}_{2n-1}} \nonumber \\
\tilde{\bold{F}}_{{M'}_1\cdots {M'}_{2n}}
  &=&\bold{G}^A_{{M'}_1\cdots{M'}_{2n}}+\bold{A}^{(1)}_{[{M'}_1}\bold{G}^A_{{M'}_2\cdots{M'}_{2n}]},
\end{eqnarray}
where
\begin{eqnarray}
  \bold{G}^A_{2n-1}&=&d\bold{C}^A_{2n-2}+\hat{\bold{H}}_3\wedge \bold{C}^A_{2n-4}-\bold{F}^{(2)}_2\wedge \bold{C}^A_{2n-3} \nonumber \\
  \bold{G}^A_{2n}&=&d\bold{C}^A_{2n-1}+\hat{\bold{H}}_3\wedge \bold{C}^A_{2n-3}-\bold{F}^{(1)}_2\wedge \bold{C}^A_{2n-2},
\end{eqnarray}
where
\begin{eqnarray}
  \bold{F}_2^{(i)}&=&d\bold{A}^{(i)}  \quad (i=1, 2) \nonumber \\
  \hat{\bold{H}}_3&=&d \bold{B}+\frac{1}{2}\bold{A}^{(1)} \wedge  \bold{F}_2^{(2)}+\frac{1}{2}\bold{A}^{(2)} \wedge  \bold{F}_2^{(1)}.
\end{eqnarray}
As a result, 
\begin{align}
  |\tilde{\bold{F}}_{2n}|^2
  =|\bold{G}^A_{2n}|^2+\bold{k}^{-2}|\bold{G}^A_{2n-1}|^2.
\end{align}
On the other hand, the $2n-1$-form field strengths follow
\begin{align}
  \tilde{\bold{F}}_{{M'}_1\cdots {M'}_{2n-2}9}
 & =-\bold{G}^B_{{M'}_1\cdots{M'}_{2n-2}} \nonumber \\
  \tilde{\bold{F}}_{{M'}_1\cdots {M'}_{2n-1}}
  &=\bold{G}^B_{{M'}_1\cdots{M'}_{2n-1}}+\bold{A}^{(1)}_{[{M'}_1}\bold{G}^B_{{M'}_2\cdots{M'}_{2n-1}]},
\end{align}
where
\begin{align}
  \bold{G}^B_{2n}&=d\bold{C}^B_{2n-1}+\hat{\bold{H}}_3\wedge \bold{C}^B_{2n-3}-\bold{F}^{(2)}_2\wedge \bold{C}^B_{2n-2} \nonumber \\
  \bold{G}^B_{2n-1}&=d\bold{C}^B_{2n-2}+\hat{\bold{H}}_3\wedge \bold{C}^B_{2n-4}-\bold{F}^{(1)}_2\wedge \bold{C}^B_{2n-3}.
\end{align}
As a result, 
\begin{align}
  |\tilde{\bold{F}}_{2n-1}|^2
  =|\bold{G}^B_{2n-1}|^2+\bold{k}^{-2}|\bold{G}^B_{2n-2}|^2.
\end{align}
To summarize, (\ref{action of T-symmetry}) under (\ref{DimensinalReduction}) becomes 
\begin{align}
  S''=&\int \mathcal{D}\bold{E} \mathcal{D}\bar{\tau} \mathcal{D}\bold{X}_{\hat{D}_T} \sqrt{-\bold{G}}\left(e^{-2\bold{\phi}}\left(\bold{R}+4(\partial_{M'}\bold{\phi})^2-4(\partial_{M'}\ln \bold{k})^2-\frac{1}{2}|\hat{\bold{H}}_3|^2+\frac{1}{2}\bold{k}^2|\bold{F}^{(1)}_2|^2+\frac{1}{2}\bold{k}^{-2}|\bold{F}^{(2)}_2|^2\right)\right. \nonumber \\
  &\left.-\frac{1}{2}\sum_{p=1}^{\infty}(\bold{k}|\bold{G}^A_{2p}|^2+\bold{k}^{-1}|\bold{G}^B_{2p}|^2+\bold{k}^{-1}|\bold{G}^A_{2p-1}|^2+\bold{k}|\bold{G}^B_{2p-1}|^2)\right). \label{lastaction}
\end{align}
This action is invariant under a $Z_2$ transformation, 
\begin{equation}
(\bold{k}, \bold{A}^{(1)}, \bold{C}_n^A) \leftrightarrow (\bold{k}^{-1}, \bold{A}^{(2)}, \bold{C}_n^B). \label{lasttrans}
\end{equation}

In \cite{Honda:2021rcd},  it is shown that 
string geometry theory is consistently truncated to type IIA and IIB supergravities by particular configurations, so called IIA and IIB string background configurations,  respectively. Namely, the equations of motion of string geometry theory substituted these configurations into are equivalent to the equations of motion of type IIA and IIB supergravities, respectively. An explicit form of the configurations are as follows:
\begin{description}
\item{metric:}
\begin{eqnarray}
&&\mathbf{G}_{00} \left(\bar{\tau}, \bold{X}_{\hat{D}_T} \right) = -1 \nonumber \\ 
&&\mathbf{G}_{(\mu_{1} \bar{\sigma}_{1} \bar{\theta}_1)(\mu_{2} \bar{\sigma}_{2} \bar{\theta}_2)}\left(\bar{\tau}, \bold{X}_{\hat{D}_T} \right)
=
G_{\mu_{1} \mu_{2}} \left( \bold{X}_{\hat{D}_T}(\bar{\sigma}_{1} \bar{\theta}_1) \right) \delta_{(\bar{\sigma}_{1} \bar{\theta}_1) (\bar{\sigma}_{2} \bar{\theta}_2)  }\delta_{(\bar{\sigma}_{1} \bar{\theta}_1) (\bar{\sigma}_{1} \bar{\theta}_1) }  
\nonumber \\
&&\text{the others} = 0,  \label{Gansatz}
\end{eqnarray}
\item{scalar field:}
\begin{eqnarray}
\bold{\Phi}\left(\bar{\tau},\bold{X}_{\hat{D}_T}\right)=\int d \bar{\sigma} 
d^2 \bar{\theta} \hat{\bold{E}} \delta_{(\bar{\sigma}, \bar{\theta}) (\bar{\sigma},\bar{\theta})}
\, \Phi(\bold{X}_{\hat{D}_T} (\bar{\sigma}, \bar{\theta})),
\label{Sansatz}
\end{eqnarray}
\item{B field:}
\begin{eqnarray}
&&\mathbf{B}_{(\mu_{1} \bar{\sigma}_{1} \bar{\theta}_1)(\mu_{2} \bar{\sigma}_{2} \bar{\theta}_2)}\left(\bar{\tau}, \bold{X}_{\hat{D}_T} \right)
=
B_{\mu_{1} \mu_{2}} \left( \bold{X}_{\hat{D}_T}(\bar{\sigma}_{1} \bar{\theta}_1) \right) \delta_{(\bar{\sigma}_{1} \bar{\theta}_1) (\bar{\sigma}_{2} \bar{\theta}_2)}\delta_{(\bar{\sigma}_{1} \bar{\theta}_1) (\bar{\sigma}_{1} \bar{\theta}_1) }\nonumber \\
&&\text{the others} = 0,  \label{Bansatz}
\end{eqnarray}
\item{p-form field:}
\begin{eqnarray}
&&\mathbf{C}_{(\mu_{1} \bar{\sigma}_{1} \bar{\theta}_1 )\cdots(\mu_{p} \bar{\sigma}_{p} \bar{\theta}_p)}\left(\bar{\tau}, \bold{X}_{\hat{D}_T} \right) = C_{\mu_{1} \cdots \mu_{p}} \left( \bold{X}_{\hat{D}_T}(\bar{\sigma}_{1} \bar{\theta}_1) \right) \delta_{(\bar{\sigma}_{1} \bar{\theta}_1) (\bar{\sigma}_{2} \bar{\theta}_2)}
\cdots
\delta_{(\bar{\sigma}_{p-1} \bar{\theta}_{p-1}) (\bar{\sigma}_{p} \bar{\theta}_p)}
\delta_{(\bar{\sigma}_{1} \bar{\theta}_1) (\bar{\sigma}_{1} \bar{\theta}_1)}
\nonumber \\
&& (p=0, 1, \cdots, 8) \nonumber \\
&&\text{the others} = 0, \label{pansatz}
\end{eqnarray}
\end{description}
where
\begin{eqnarray}
&&C_{\mu_{1} \cdots \mu_{p}}=0 \mbox{  (p: even)}
\nonumber \\
&&\tilde{F}_8=-* \tilde{F}_2
\nonumber \\
&&\tilde{F}_6=* \tilde{F}_4,
\label{IIAconfiguration}
\end{eqnarray}
for IIA string background configuration, or
\begin{eqnarray}
&&C_{\mu_{1} \cdots \mu_{p}}=0 \mbox{  (p: odd)}
\nonumber \\
&&\tilde{F}_9=* \tilde{F}_1
\nonumber \\
&&\tilde{F}_7=-* \tilde{F}_3
\nonumber \\
&&\tilde{F}_5=* \tilde{F}_5,
\label{IIBconfiguration}
\end{eqnarray}
for IIB string background configuration. If these configurations are substituted to the equations of motion of (\ref{action of T-symmetry}), we obtain the equations of motions of 
\begin{eqnarray}
S_{IIA}& =& \frac{1}{2 \kappa^{2}_{10}} \Biggl( \int d^{10}x \sqrt{-G}  \left( e^{ -2 \phi } \left( R + 4 \nabla_{\mu} \phi \nabla^{\mu} \phi - \frac{1}{2} |H|^{2}  \right)
- \frac{1}{2} |\tilde{F}_2|^{2}- \frac{1}{2} |\tilde{F}_4|^{2} \right) \nonumber \\
&&-\frac{1}{2}\int B \wedge dC_3 \wedge dC_3
\Biggr),
\label{IIAaction}
\end{eqnarray}
in the IIA case \cite{Huq:1983im, Giani:1984wc, Campbell:1984zc}, and equations of motions of
\begin{eqnarray}
S_{IIB}& =& \frac{1}{2 \kappa^{2}_{10}} \Biggl( \int d^{10}x \sqrt{-G}  \left( e^{ -2 \phi } \left( R + 4 \nabla_{\mu} \phi \nabla^{\mu} \phi - \frac{1}{2} |H|^{2}  \right)
- \frac{1}{2} |\tilde{F}_1|^{2}- \frac{1}{2} |\tilde{F}_3|^{2} - \frac{1}{4} |\tilde{F}_5|^{2} \right) \nonumber \\
&&-\frac{1}{2}\int C_4 \wedge H \wedge dC_2
\Biggr),
\label{IIBaction}
\end{eqnarray}
and the self-dual condition,
\begin{eqnarray}
\tilde{F}_5=* \tilde{F}_5,
\nonumber
\end{eqnarray}
in the IIB case \cite{Schwarz:1983qr, Howe:1983sra}. That is, the string geometry theory  (\ref{action of T-symmetry}) is consistently truncated to type IIA and IIB supergravities (\ref{IIAaction}) and (\ref{IIBaction}), respectively \cite{Honda:2021rcd}. By imposing the condition (\ref{DimensinalReduction}) on the string background configurations,
we obtain 
\begin{equation}
\partial_9 \varphi=0, \label{SUGRAcondition}
\end{equation}
for  all the supergravity fields $\varphi$.
 By decomposing the string background configurations (\ref{Gansatz})  $\sim$ (\ref{IIBconfiguration})  as in (\ref{decomposition1}) $\sim$ (\ref{decomposition2}), the fields of the supergravities are decomposed as
\begin{align}
  G_{\mu \nu}=\left(
    \begin{array}{cc}
      g_{{\mu'}{\nu'}}+k^2 A^{(1)}_{\mu'} A^{(1)}_{\nu'} & -k^2 A^{(1)}_{\mu'} \\
      -k^2 A^{(1)}_{\nu'} & k^2
    \end{array}
  \right),
\end{align}
where
$\mu'=0, 1, \cdots, 8,$
\begin{equation}
\Phi= \phi+\frac{1}{2}\ln k
\end{equation}
\begin{align}
  B^{[10]}_{{\mu'} 9}&=A^{(2)}_{\mu'} \\
  B^{[10]}_{{\mu'}{\nu'}}&=B_{{\mu'}{\nu'}}+A^{(1)}_{[{\mu'}}A^{(2)}_{{\nu'}]},
\end{align}
and
\begin{align}
  C^{[10]}_{{\mu'}_1\cdots{\mu'}_{2n-2}9}&=C_{{\mu'}_1\cdots{\mu'}_{2n-2}} \label{eq:cmdB} \\
  C^{[10]}_{{\mu'}_1\cdots{\mu'}_{2n-1}}&= C_{{\mu'}_1\cdots{\mu'}_{2n-1}}-(2n-1)A^{(1)}_{[{\mu'}_1}C_{{\mu'}_2\cdots{\mu'}_{2n-1    }]}, \label{eq:cmnB}
\end{align}
in the IIA case and
\begin{align}
  C^{[10]}_{{\mu'}_1\cdots{\mu'}_{2n-1}9}&= -C_{{\mu'}_1\cdots{\mu'}_{2n-1}} \label{eq:cmd} \\
  C^{[10]}_{{\mu'}_1\cdots{\mu'}_{2n}}&= C_{{\mu'}_1\cdots{\mu'}_{2n}}-2nA^{(1)}_{[{\mu'}_1}C_{{\mu'}_2\cdots{\mu'}_{2n}]}, \label{eq:cmn} 
\end{align}
in the IIB case, where $[10]$s in the left hand sides represent the fields before the decomposition.
We decompose all the fields other than a higher R-R forms (higher than the 4 form) in this way.  The decomposition of the higher R-R forms are determined by the Poincar$\acute{\mbox{e}}$ dual relation (\ref{IIAconfiguration}) and  (\ref{IIBconfiguration}).

By  imposing the condition (\ref{SUGRAcondition}) and substituting these decompositions, the type  IIA supergravity action (\ref{IIAaction}) and the type IIB one (\ref{IIBaction})  are transformed into each other, under 
\begin{equation}
(k, A^{(1)}) \leftrightarrow (k^{-1}, A^{(2)}).  \label{trans2}
\end{equation}
This relation is so-called T-duality between the type IIA and IIB supergravities \cite{Buscher:1987sk, Ortin}, where  the transformation is called  the Buscher rule  \cite{Buscher:1987sk} if it is expressed as a transformaion of the fields before the decomposition. Therefore,  the invariance of the action (\ref{lastaction}) under the transformation (\ref{lasttrans}) should be called T-symmetry in string geometry theory\footnote{If we name $C_n$ $C^A_n$ in the right hand side of (\ref{eq:cmdB}) and (\ref{eq:cmnB}) and name $C_n$ $C^B_n$ in the right hand side of (\ref{eq:cmd}) and (\ref{eq:cmn}), the transformation (\ref{trans2}) becomes $(k, A^{(1)}, C^A_n) \leftrightarrow (k^{-1}, A^{(2)}, C^B_n) $ and the relation to the transformation (\ref{lasttrans}) becomes manifest.} . 

In \cite{Sato:2022brv, Nagasaki:2023fnz}, the path-integrals of the type IIA and IIB perturbative strings are shown to be obtained from the two-point correlation functions of the quantum parts of $G_{10\, 10}$ around  the perturbative vacua, which consist of the classical parts of $G_{10 \, 10}$ fixed to vacuum configurations  and the type IIA and IIB string background configurations  (\ref{Gansatz})  $\sim$ (\ref{IIBconfiguration}), respectively. 
$G_{10 \, 10}$ is invariant under the T-symmetry transformation (\ref{lasttrans}), where the type IIA and IIB string background configurations are exchanged to each other. 
Thus, the path-integrals of the type IIA and IIB perturbative strings are exchanged under the T-symmetry transformation if the transformation of the quantum parts is defined so that  $G_{10 \, 10}$ is invariant against the transformation of the classical parts.   The spectrums and the scattering amplitudes of  the strings, which are obtained from the path-integrals of the type IIA and IIB pertrubative strings, correspond to each other under the T-symmetry transformation, where the Kalzua Klein modes in the type IIA strings and the winding modes in the type IIB strings are exchanged and vice versa, and the massive modes are also exchanged. Therefore, the T-symmetry transformation (\ref{lasttrans}) is also the T-dual transformation itself at the level of the perturbative strings beyond the supergravity.     

\subsection{T-symmetry in the temporal direction}

In this subsection, we will show that there is a symmetry independent of the T-duality between the perturbative string theories, by performing a dimensional reduction in $\bar{\tau}=X^{10}$ direction on the action (\ref{action of T-symmetry}), where all the fields $\boldsymbol{\varphi}$ satisfy 
\begin{equation}
\partial_{10} \boldsymbol{\varphi}=0.  \label{condition}
\end{equation} 
We decompose the metric as
\begin{align}
  \bold{G}_{\bold{M} \bold{N}}=\left(
    \begin{array}{cc}
      \bold{g}_{{M'}{N'}}-\bold{k}^2 \bold{A}^{(1)}_{M'} \bold{A}^{(1)}_{N'} & +\bold{k}^2 \bold{A}^{(1)}_{M'} \\
      +\bold{k}^2 \bold{A}^{(1)}_{N'} & -\bold{k}^2
    \end{array}
  \right),
\end{align}
where $M'=(\mu \bar{\sigma} \bar{\theta})$ and $N'=(\mu' \bar{\sigma}' \bar{\theta}')$, and the other fields as in (\ref{dil}) $\sim$ (\ref{decomposition2}) where we change the 9-direction to the 10-direction. 
As a result, (\ref{action of T-symmetry}) under (\ref{condition}) becomes 
\begin{align}
  S''=&\int \mathcal{D}\bold{E}\mathcal{D}\bold{X}_{\hat{D}_T} \sqrt{\bold{G}}\left(e^{-2\bold{\phi}}\left(\bold{R}+4(\partial_{M'}\bold{\phi})^2-4(\partial_{M'}\ln |\bold{k}|)^2-\frac{1}{2}|\hat{\bold{H}}_3|^2+\frac{1}{2}\bold{k}^2|\bold{F}^{(1)}_2|^2+\frac{1}{2}\bold{k}^{-2}|\bold{F}^{(2)}_2|^2\right)\right. \nonumber \\
  &\left.-\frac{1}{2}\sum_{p=1}^{\infty}(\bold{k}|\bold{G}^A_{2p}|^2-\bold{k}^{-1}|\bold{G}^B_{2p}|^2-\bold{k}^{-1}|\bold{G}^A_{2p-1}|^2+\bold{k}|\bold{G}^B_{2p-1}|^2)\right).
\label{truefinalaction}
\end{align}
This action is invariant under a $Z_2$ transformation, 
\begin{equation}
(\bold{k}, \bold{A}^{(1)}, \bold{C}_n^A) \leftrightarrow (-\bold{k}^{-1}, \bold{A}^{(2)}, \bold{C}_n^B).  \label{TheLastTransform}
\end{equation}
Under the string background configurations (\ref{Gansatz})  $\sim$ (\ref{IIBconfiguration}), this decomposition results in  $\bold{k}^2=1$ and $\bold{A}^{(1)}=\bold{A}^{(2)}=0$. Then, under the T-symmetry transformation (\ref{TheLastTransform}), the metric and B-field are not transformed in the supergravities\footnote{The R-R fields in the IIA (IIB) supergravity transform not to the fields in the IIB (IIA) supergravity, but to other parts of the string geometry fields.}. Therefore, the T-symmetry transformation in this case is independent of the T-dual transformation (\ref{trans2}), and gives the symmetry that cannot be seen in the perturbative string theories.








\vspace{1cm}

\section{Conclusion and discussion}
\setcounter{equation}{0}
The fields of the string geometry theory (\ref{action of bos string-geometric model}) include the fields of the ten-dimensional supergravities, which represent string backgrounds. By consistently truncating the string geometry fields to the supergravity fields, one can obtain all the type, namely Type IIA, IIB, SO(32) type I, and SO(32) and $E_8 \times E_8$ heterotic  ten-dimensional supergravities from the string geometry theory  \cite{Honda:2021rcd}. In this paper,  we have shown that the dimensionally reduced string geometry action (\ref{lastaction}) is invariant under what we call T-symmetry. In case of the dimensional reduction in $\bold{X}_{\hat{D}_T}^{(9 \bar{\sigma} \bar{\theta})}$ directions, the T-symmetry transformation acts on the IIA and IIB supergravity fields in the string geometry fields as  the T-dual transformation, and thus the IIA and IIB supergravities that are obtained by the consistent truncations,  are transformed to each other. 
Moreover, as we discussed in section 3, the T-symmetry transformation (\ref{lasttrans}) is also  the T-dual transformation itself at the level of the perturbative strings beyond the supergravity, because      
the fluctuations around the type IIA and IIB perturbative vacua are exchanged into each other under the T-symmetry transformation, and the fluctuations give the path-integrals of the type IIA and IIB perturbative strings, respectively \cite{Sato:2022brv, Nagasaki:2023fnz}.
That is, the T-symmetry transformation gives the T-dual transformation between the perturbative vacua. 
 In case of  the dimensional reduction in  $\bar{\tau}$ direction,  the T-symmetry transformation is independent of the T-dual transformation, and gives the symmetry that cannot be seen in the perturbative string theories. This symmetry will be worked out in detail in near future.

In this way, T-symmetry constrains actions strongly because it is highly non-trivial. The action of string geometry theory that includes terms up to two-derivatives, would be determined as (\ref{action of bos string-geometric model}), because  all the ten-dimensional supergravities are derived by consistent truncations from it, and its dimensionally reduced action (\ref{lastaction}) has T-symmetry.

So-called exotic branes  \cite{deBoer:2010ud,Bergshoeff:2011se,deBoer:2012ma,Sakatani:2014hba}  are conjectured to exist  in string theory  in consistency with T-duality. However, for example, $5^2_2$ brane  \cite{Kikuchi:2012za,Hassler:2013wsa,Geissbuhler:2013uka,Kimura:2013zva,Chatzistavrakidis:2013jqa} cannot exist in supergravity  because it is multi-valued as an ordinary manifold.
On the other hand in \cite{Honda:2021rcd}, the string geometry theory (\ref{action of bos string-geometric model}) is shown to be consistently truncated to a ten-dimensional gravity \cite{Fukuma:1999jt} that has O(d,d) symmetry if it is compactified on $T^d$ and can be consistently truncated to the bosonic part of the type IIA and IIB supergravities.   
 In this ten-dimensional gravity, $5^2_2$ brane can exist because the multi-valuedness  can be absorbed by the O(d,d) transformation. Therefore, an exotic brane exists consistently in string geometry theory. That is, string geometry includes a class of geometry consistent with T-duality beyond ordinary manifolds.

T-duality will become T-symmetry in a non-pertubatively formulated string theory because it includes both type IIA and IIB perturbative vacua, which are T-dual to each other. This apparently contradicts the conjecture that  quantum gravity does not have a global symmetry, including a discrete symmetry \cite{Banks:1988yz, Vafa:2005ui, Arkani-Hamed:2006emk, Harlow:2018tng}. Although T-symmetry in string geometry theory is a  global discrete symmetry, it is not a complete one, but an effective one, namely a symmetry when the theory is dimensionally reduced. As a result, string geometry theory resolves the paradox between T-duality and quantum gravity.

\vspace*{0cm}

\section*{Acknowledgement}
We would like to thank  
M. Fukuma,
Y. Sakatani,
S. Sasaki,
K. Shiozawa,
Y. Sugimoto,
and especially 
H. Kawai, A. Tsuchiya, and T. Yoneya
for long and valuable discussions. 
The work of M. S. is supported by Hirosaki University Priority Research Grant for Future Innovation.

\end{document}